\documentclass[12pt]{article}

\usepackage[a4paper, margin=1in]{geometry}
\usepackage{setspace}        
\usepackage{amssymb}
\usepackage{mathtools}    
\usepackage{amsthm}
\usepackage{graphicx}
\usepackage{pgfplots}
\pgfplotsset{compat=1.17}
\usepackage{booktabs}
\usepackage{multirow}
\usepackage{enumitem}
\usepackage{caption}
\usepackage{authblk}
\usepackage[authoryear]{natbib}

\usepackage{hyperref}

\title{Lasso--Ridge Refitting: A Two-Stage Estimator for High-Dimensional Linear Regression}
\author[1]{Guo Liu}
\affil[1]{Department of Pure and Applied Mathematics, Waseda University}

\theoremstyle{plain}
\newtheorem{theorem}{Theorem}[section]
\newtheorem{lemma}[theorem]{Lemma}
\newtheorem{proposition}[theorem]{Proposition}
\newtheorem{corollary}[theorem]{Corollary}
\theoremstyle{definition}
\newtheorem{definition}[theorem]{Definition}
\newtheorem{remark}[theorem]{Remark}
\newtheorem{assumption}[theorem]{Assumption}
\begin{document}
	\pagenumbering{gobble}  
	\maketitle
	\renewcommand{\thesection}{\arabic{section}}
	\pagenumbering{roman}
	  
	\section*{Abstract}
	\addcontentsline{toc}{chapter}{Abstract}
	The least absolute shrinkage and selection operator (Lasso) is a popular method for high-dimensional statistics. However, it is known that the Lasso often has estimation bias and prediction error. To address such disadvantages, many alternatives and refitting strategies have been proposed and studied. This work introduces a novel Lasso--Ridge method. Our analysis indicates that the proposed estimator achieves improved prediction performance in a range of settings, including cases where the Lasso is tuned at its theoretical optimal rate \( \sqrt{\log(p)/n}\). Moreover, the proposed method retains several key advantages of the Lasso, such as prediction consistency and reliable variable selection under mild conditions. Through extensive simulations, we further demonstrate that our estimator  outperforms the Lasso in both prediction and estimation accuracy, highlighting its potential as a powerful tool for high-dimensional linear regression.

	\setstretch{1} 
	\pagenumbering{arabic} 
	
	\section{Introduction }
	High-dimensional linear regression, where the number of predictors \(p\) exceeds the number of observations \(n\), commonly arises in a variety of scientific fields, including genomics, signal processing, and finance. In such settings, the goal is often to identify a sparse set of relevant variables while achieving good predictive and estimation performance. The Lasso is a widely used method for high-dimensional regression \citep{tibshirani1996regression} that performs simultaneous variable selection and estimation by imposing an \(\ell_1\) penalty.

The theoretical properties of the Lasso have been extensively studied. Under suitable conditions, the Lasso satisfies oracle inequalities and achieves estimation and prediction consistency \citep{geer2016estimation}. However, a well-known drawback of the Lasso is that it introduces bias due to the \(\ell_1\)-penalty. Moreover, it has been shown that there exists a lower bound on the prediction error \citep{JMLR:v19:17-025}, and that the Lasso with a single choice of the penalty parameter cannot simultaneously achieve variable selection consistency and \(\sqrt{n}\)-consistency \citep{lahiri2021necessary}.

To address these drawbacks and achieve better performance, a variety of refitting strategies have been developed \citep{chzhen2019lasso}. These methods typically aim at modifying the initial Lasso estimate to reduce shrinkage bias and enhance both prediction and estimation accuracy. For instance, the OLS post-Lasso \citep{belloni2013least} applies ordinary least squares to the Lasso-selected support to reduce bias. The Adaptive Lasso \citep{zou2006adaptive} uses a reweighted \(\ell_1\) penalty in a second-stage Lasso to better reflect the relative sizes of the coefficients. The Relaxed Lasso \citep{meinshausen2007relaxed} adjusts the degree of penalization after variable selection, allowing greater flexibility in balancing sparsity and shrinkage. These approaches reflect a general two-stage refitting strategy: The Lasso is first applied to obtain an initial result, such as variable selection, and then a second step is carried out based on that result. These methods have been studied both theoretically and empirically, highlighting their advantages in various settings. However, most existing work focuses on consistency and empirical studies, while few results provide direct comparisons or probability inequalities that demonstrate prediction improvement.

In this paper, we propose a novel Lasso--Ridge refitting approach. The intuition behind our method is to fill in the prediction gap introduced by the Lasso's penalty by applying a ridge regression \citep{hoerl1970ridge} that brings a safe correction targeting the active set of the Lasso estimator. The main contributions of this work can be summarized as follows:
\begin{enumerate}
	\item We establish theoretical guarantees for improved prediction performance under mild conditions on the design matrix, noise distribution, and the Lasso tuning parameter.
	\item We show that the proposed method preserves the sign pattern selected by the Lasso and inherits its prediction consistency.
	\item We demonstrate improved performance in both prediction and estimation across a diverse set of numerical studies.
\end{enumerate}

	\section{Preliminaries and Background }
	\subsection{Notation, Norms, and Model Assumptions}
We use \( \|\cdot\|_1 \) and \( \|\cdot\|_2 \) to denote the \( \ell_1 \) and Euclidean norms, respectively. For a matrix \( A \in \mathbb{R}^{n \times p} \), we denote the operator \(\infty\)-norm by
\[
\|A\|_\infty := \max_{1 \leq i \leq n} \sum_{j=1}^p |A_{ij}|,
\]
which corresponds to the maximum absolute row sum. The spectral norm, denoted \( \|A\|_2 \), is the largest singular value of \( A \) and coincides with the operator norm induced by the Euclidean norm:
\[
\|A\|_2 := \sup_{\|x\|_2 = 1} \|Ax\|_2.
\]
For a finite set \(E\), we denote its cardinality by \(|E|\). For any index set \( S \subseteq [p] := \{1, \dots, p\} \), we write \( A_S \in \mathbb{R}^{n \times |S|} \) to denote the submatrix of the design matrix \( A \in \mathbb{R}^{n \times p} \) consisting of the columns indexed by \( S \). That is, \( A_S \) is formed by extracting the columns \( A_{\cdot j} \) for each \( j \in S \). Similarly, for a vector \( \beta \in \mathbb{R}^p \), we write \( \beta_S \in \mathbb{R}^{|S|} \) for the subvector containing the entries indexed by \( S \).
For completeness, we adopt standard conventions in high-dimensional statistics to ensure that all expressions remain well-defined, even in degenerate cases. For example, when \( S = \emptyset \), we interpret \( A_S \in \mathbb{R}^{n \times 0} \). In our proofs, we explicitly separate cases involving empty index sets to simplify the argument structure. Throughout, care is taken to maintain mathematical rigor, and no ambiguity arises in definitions or proofs.

We consider a standard linear regression model given by
\[
y = X\beta^0 + \varepsilon, \, \,\,\varepsilon\sim N(0,\sigma^2I_n),
\]
where \(y = (y_1, \ldots, y_n) \in \mathbb{R}^n\) is the response vector, \(X \in \mathbb{R}^{n \times p}\) is the design matrix, \(\beta^0\) denotes the unknown coefficient vector, and \(\varepsilon\) is the noise vector. This Gaussian assumption is made for simplicity, and all theoretical results below also hold when \( \varepsilon \) is a sub-Gaussian random vector with mean zero and sub-Gaussian parameter \( \sigma \). We additionally assume that the columns of \(X\) are normalized in such a way that for all \(j \in [p]\), we have \(\|X_j\|_2^2 = n\), where \(X_j\) represents the \(j\)th column of the matrix \(X\). 
\subsection{The Lasso Estimator and Its Basic Properties}
The Lasso estimator is noted as \(\hat{\beta}_{\mathrm{L}}\), with its definition as follows:
\[
\hat{\beta}_{\mathrm{L}}=\underset{\beta \in \mathbb{R}^{p}}{\arg\min} \left\{ \frac{1}{2n} \| y - X\beta \|_2^2 +  \lambda_{\mathrm{L}}\|\beta \|_1 \right\}.
\]
\begin{definition}
	We define the key quantities associated with the Lasso:
	\begin{itemize}
		\item \emph{The equicorrelation set:}
		\[
		E \coloneqq \{\, i \in \{1,2,\ldots,p\} : |\frac{1}{n}X_i^\top (y - X\hat{\beta}_{\mathrm{L}})| = \lambda_{\mathrm{L}} \,\}.
		\]
		\item \emph{Restricted Gram matrix:}
		\[
		\Sigma_{n,E} \coloneqq \frac{X_{E}^{\top} X_{E}}{n}.
		\]
		\item \emph{The equicorrelation signs:}
		\[
			s \coloneqq \operatorname{sign}({X_E}^T(y - X\hat{\beta}_{\mathrm{L}}))
		\]
	\end{itemize}
\end{definition}
\begin{assumption}
	The columns of the design matrix \( X \) are in general position.
\end{assumption}
\begin{proposition}[Uniqueness of the Lasso Solution \citep{tibshirani2013lasso}]
	Under Assumption 2.12, the solution \(\hat{\beta}_{\mathrm{L}} \) is unique. Moreover, its selected support coincides with the \(E\), to be more specific, \(\hat{\beta}_{\mathrm{L},j}\neq 0 \) if and only if \( j \in E \). 
\end{proposition}
\begin{remark}
	This is a mild condition that holds with probability one when the entries of the design matrix \( X \) are drawn from a continuous distribution. It ensures the uniqueness of Lasso solutions, thereby avoiding ambiguity when defining a refitting step based on the non-zero part of Lasso estimators. In the rest of the paper, we adopt this assumption without further mention.
	In addition, without this assumption, it also holds that every Lasso estimator \(\hat{\beta}_{\mathrm{L}}\) corresponding to a fixed tuning parameter \(\lambda_{\mathrm{L}}\) gives the same fitted value \(X\hat{\beta}_{\mathrm{L}}\), which ensures the uniqueness of the equicorrelation set \(E\) and the equicorrelation signs \(s\).
\end{remark}
\begin{proposition}
	The KKT conditions for the Lasso estimator can be derived from subdifferential calculus \citep{geer2016estimation}. Introducing \( \bar{z} \) as an element of the subdifferential of the \( \ell_1 \) norm at \( \hat{\beta}_{\mathrm{L}} \), the conditions are
	\[
	\frac{1}{n} X^\top (y - X \hat{\beta}_{\mathrm{L}}) = \lambda_{\mathrm{L}} \bar{z},
	\]
	where \( \bar{z} \in \partial \|\hat{\beta}_{\mathrm{L}}\|_1 \) and is given componentwise by:
	\[
	\bar{z}_i =
	\begin{cases}
		\mathrm{sign}(\hat{\beta}_{\mathrm{L},i}), & \text{if } \hat{\beta}_{\mathrm{L},i} \neq 0, \\
		u_i \in [-1, 1], & \text{if }\hat{\beta}_{\mathrm{L},i} = 0,
	\end{cases}
	\quad \text{for } i = 1, \dots, p.
	\]
	Also, notice that
	\[
	\frac{1}{n} X_E^\top (y - X \hat{\beta}_{\mathrm{L}}) = \lambda_{\mathrm{L}}s.
	\]
	Here, \( \mathrm{sign}(\cdot) \) denotes the sign function.
\end{proposition}

	\section{Definition of a Lasso--Ridge Estimator}
	A novel Lasso-Ridge estimator could be defined as:
\begin{align*}
	\hat{\beta}_{\mathrm{R}} =\underset{\beta \in \mathbb{R}^{p}, \beta_{-E}=0}{\arg\min} \left\{ \frac{1}{2n} \| y - X\beta \|_2^2 +  \frac{\lambda_{\mathrm{R}}}{2}\|\beta-\hat{\beta}_{\mathrm{L}} \|_2^2  \right\},
\end{align*}
or equivalently:
\begin{align*}
 \hat{\beta}_{\mathrm{R}} = \hat{\delta} + \hat{\beta}_{\mathrm{L}},
\end{align*}
\begin{align*}
\quad \hat{\delta} = \underset{\delta \in \mathbb{R}^ {p}, \delta_{-E}=0}{\arg\min} \left\{ \frac{1}{2n} \| (y - X \hat{\beta}_{\mathrm{L}})-X\delta \|_2^2 +  \frac{\lambda_{\mathrm{R}}}{2}\| \delta \|_2^2 \right\},
\end{align*}
where \(\hat{\beta}_{\mathrm{L}}\) represents the Lasso estimator, \(E\) denotes the equicorrelation set of the Lasso estimator, which coincides with the set of nonzero coefficients under Assumption~2.2. For given \(y\), Lasso selects no columns when \(\lambda_{\mathrm{L}}\) is large enough, and because \(\varepsilon\) can be any vector in \(\mathbb{R}^n\), the Lasso may yield a zero solution and select no variables. In such cases, \(\hat{ \delta}\) is a zero vector and no refitting is made to the Lasso estimator.
When \(E\) is not an empty set, \(\hat{ \delta}_E\) could be regarded as a ridge estimator:
\begin{align*}
	\quad \hat{\delta}_E = \underset{\delta \in \mathbb{R}^ {|E|}}{\arg\min} \left\{ \frac{1}{2n} \| (y - X \hat{\beta}_{\mathrm{L}})-X_E\delta_E \|_2^2 +  \frac{\lambda_{\mathrm{R}}}{2}\| \delta_E \|_2^2 \right\}.
\end{align*}

The tuning parameter \(\lambda_{\mathrm{R}}\) controls the second step. As \(\lambda_{\mathrm{R}}\) tends to infinity, the second step vanishes, and the resulting estimator reduces to the original Lasso solution. On the other hand, when \(\lambda_{\mathrm{R}}=0\), this refitting strategy is the same as least square refitting after the Lasso. In our theoretical analysis, the tuning parameter used in the second step depends on both the initial Lasso tuning parameter \( \lambda_{\mathrm{L}} \) and the noise vector \( \varepsilon \). For simplicity, we denote it simply by \( \lambda_{\mathrm{R}} \) throughout the paper.

The Lasso tuning parameter, denoted by \(\lambda_{\mathrm{L}}\), simultaneously governs variable selection and estimation. However, previous works suggest that it may be difficult for one choice of \(\lambda_{\mathrm{L}}\) to simultaneously yield reliable variable selection and strong predictive performance. In some cases, \(\lambda_{\mathrm{L}}\) is chosen conservatively to ensure sparsity, leaving room for a second-stage procedure to reduce bias or improve estimation. This refitting strategy modifies the nonzero coefficients of the Lasso estimator in order to further reduce the squared loss on the selected variables. To be more specific, we have the following basic property:
\begin{proposition}[Reduction in Empirical Risk]
	\[
	\| y - X\hat{\beta}_{\mathrm{R}} \|_2 \leq  \| y - X\hat{\beta}_{\mathrm{L}} \|_2.
	\]
\end{proposition}
Next, we establish theoretical results for cases where the proposed approach improves prediction performance.

	\section{Theoretical Results}
	\subsection{Enhanced Prediction Performance}
\begin{theorem}[Pointwise Prediction Improvement]
	Select the tuning parameter for the second step \(\lambda_{\mathrm{R}} > 2\|\Sigma_{n,E}\|_\infty\), where \(\Sigma_{n,E}=X_E^{\top}X_E/n\). Then the Lasso-Ridge estimator satisfies the following inequality:
	\[
	\frac{1}{2n}\|X\hat{\beta}_{\mathrm{R}} - X\beta_0\|_2^2 
	\leq 
	\frac{1}{2n}\|X\hat{\beta}_{\mathrm{L}} - X\beta_0\|_2^2 
	- R(\varepsilon, X, \beta_0),
	\]
	where 
	\[
	R(\varepsilon, X, \beta_0)=\frac{\lambda_{\mathrm{R}}}{2\lambda_{\mathrm{L}}}(\lambda_{\mathrm{L}}-3\|\frac{1}{n}X^\top\varepsilon\|_\infty)\|\hat{ \delta}\|_2^2.
	\]
\end{theorem}
One crucial idea behind the prediction improvement is to select the tuning parameter \(\lambda_{\mathrm{R}} > 2\|\Sigma_{n,E}\|_\infty\) based on the support \(X_E\) obtained from the Lasso step. It is worth noting that when the number of selected columns is small or the selected columns exhibit low correlation, a strong penalty in the second step is unnecessary, allowing greater flexibility for the second step. It is reasonable to consider cases where the support selected by Lasso is close to or smaller than the true support, which implies that the penalty in the second step need not be too large in order to preserve this property. 

A lower bound on \(\|\hat{\delta}\|_{2}\) can be obtained by 
applying an orthogonal decomposition argument based on Lemma~A.4.  
This bound guarantees that the improvement achieved by the proposed 
refinement is not negligible for a suitable range of values of 
\(\lambda_{\mathrm{L}}\). Since this result is not
required for the subsequent analysis, we omit the explicit bound here.

While this tuning choice facilitates theoretical analysis, in practice there is more flexibility in selecting the penalty, and more reliable methods such as cross-validation can be used. In certain extreme cases where the Lasso tuning parameter \( \lambda_{\mathrm{L}} \) is not set at an optimal rate for prediction or estimation, such as when \( \lambda_{\mathrm{L}} \gg \sqrt{\log (p) / n}\) , substantial improvements in prediction accuracy can be observed, and this situation may occur when \( \lambda_{\mathrm{L}} \) is chosen solely to ensure variable selection consistency \citep{lahiri2021necessary}. From this theorem, we can straightforwardly derive the following corollaries, which analyze the case where \( \lambda_{\mathrm{L}} \) is chosen according to theoretical guidelines for favorable Lasso performance.
\begin{corollary}[Probability of Prediction Superiority under Fixed Design]
	Select the tuning parameter for the second step \(\lambda_{\mathrm{R}} > 2\|\Sigma_{n,E}\|_\infty\). Consider the case where \( \lambda_{\mathrm{L}}=3 \sigma \sqrt{2\log(2p/\alpha)/n}\) for some  \( \alpha\in (0, 1)\) , we have
	\begin{align*}
		P(\|X\beta^0 - X\hat{\beta}_{\mathrm{R}}\|_2\leq\| X\beta^0 - X\hat{\beta}_{\mathrm{L}} \|_2)&\geq 1 - \alpha.\\
	\end{align*}
\end{corollary}
\begin{corollary}[Asymptotic Prediction Improvement]
Select the tuning parameter for the second step \(\lambda_{\mathrm{R}} > 2\|\Sigma_{n,E}\|_\infty\). For \( \lambda_{\mathrm{L}}=C \sigma \sqrt{\log(p)/n} \) where \(C\) is a universal constant, we have
\begin{align*}
	P(\|X\beta^0 - X\hat{\beta}_{\mathrm{R}}\|_2\leq\| X\beta^0 - X\hat{\beta}_{\mathrm{L}} \|_2)\to0\quad\text{as } n \to \infty.\\
\end{align*}
\end{corollary}
\begin{remark}
	In Lasso theory, the condition \( \lambda_{\mathrm{L}} \geq \|X^\top \varepsilon / n\|_\infty \) is standard for establishing error bounds. Since \(\|X^\top \varepsilon / n\|_\infty\) is of order \(\sqrt{\log(p)/n}\), one typically chooses \( \lambda_{\mathrm{L}} \) of order \(\sqrt{\log(p)/n}\), which ensures the condition holds with high probability. From these considerations, the cases examined in our theorem correspond to the optimally tuned Lasso setting. 
\end{remark}
\subsection{Inheritance of Consistency Properties}
One might reasonably question whether the proposed Lasso--Ridge refitting procedure is simply designed to take advantage of specific structural features of the Lasso solution, potentially at the expense of its well-known consistency properties. However, as the following results demonstrate, this refitting approach retains the key consistency properties of Lasso. In this sense, the method may be regarded as a refinement of the Lasso that makes use of its strengths while modestly addressing some of its limitations.
\begin{proposition}[Sign Preservation]
	If \(\lambda_{\mathrm{R}} > 2\|\Sigma_{n,E}\|_\infty\), we have:
	\[
	\operatorname{sign}(\hat{\beta}_{\mathrm{R}}) = \operatorname{sign}(\hat{\beta}_{\mathrm{L}}).
	\]
\end{proposition}
This proposition guarantees that, when the second tuning parameter is suitably chosen, the proposed method preserves the sign pattern of the Lasso estimator. Previous work has further established that refitting procedures with this property enjoy prediction guarantees under the Restricted Eigenvalue condition \citep{bickel2009simultaneous}; see \citep{chzhen2019lasso} for further details. For a detailed proof of this proposition, see Lemma~A.5 in the appendix. Finally, we show that the proposed estimator inherits part of the prediction consistency enjoyed by the Lasso.
\begin{proposition}[Weak Prediction Consistency]
	If \( \lambda_{\mathrm{L}} \asymp \sigma \sqrt{ {\log p}/{n} }  \) and \( \lambda_{\mathrm{R}} > 2\|\Sigma_{n,E}\|_\infty\), we have:
	\[
	\frac{1}{n} \| X\hat{\beta}_{\mathrm{R}} - X\beta^0 \|_2^2 = O_p\left( \sigma \|\beta^0\|_1 \sqrt{ \frac{\log p}{n} } \right).
	\]
\end{proposition}

	\section{Numerical Studies}
	\subsection{Simulation Experiments}
We adopt a simulation framework similar to that used in the Relaxed Lasso experiments \citep{meinshausen2007relaxed} to compare the performance of our estimator with that of the Lasso across diverse settings. The response variable follows the linear model:
\[
y = X \beta + \varepsilon,
\]
where the design matrix \( X \in \mathbb{R}^{n \times p} \) consists of rows independently drawn from a multivariate normal distribution with zero mean and covariance matrix \(\Sigma\), given by
\[
\Sigma_{ij} = \rho^{|i-j|}, \quad \text{with } \rho \in \{0, 0.5\}.
\]
The noise vector \( \varepsilon \in \mathbb{R}^n \) follows a normal distribution \( \mathcal{N}(0, \sigma^2 I_n) \), with  \(\sigma \in \{ 0.5,1\}\) representing different noise level. The response vector \( y \) is then generated according to the linear model above. We consider sample sizes \(n \in \{50, 100, 200\}\) and numbers of predictors \(p \in \{100, 200, 400\}\), and the number of relevant (nonzero) coefficients is \( s \in \{5, 10, 30\} \). For \( k \leq s \), we set \( \beta_k = 1 \), and for \( k > s \), we set \( \beta_k = 0 \).

For the Lasso method, the tuning parameter \( \lambda_{\mathrm{L}} \) is selected by cross-validation from 20 logarithmically spaced values ranging from \(10^{-3}\|X^\top y\|_\infty / n\) to \(\|X^\top y\|_\infty / n\). For the Lasso–Ridge method, the search grid expands to \(20 \times 10\) candidates, where the second parameter is selected from a logarithmically spaced sequence ranging from \(10^{-3}n\) to \(n\). All estimators use 5-fold cross-validation to select the regularization parameters. Each simulation setting is repeated 100 times, and the results are averaged over these repetitions.

Performance is evaluated in terms of both prediction and estimation accuracy. In particular, we report the relative improvement of the Lasso--Ridge method over the standard Lasso, defined as
\[
\text{Improvement} \;=\; 100 \cdot \left( \frac{\overline{MSE}_{\text{Lasso}}}{\overline{MSE}_{\text{New}}} - 1 \right),
\]
where \( L_{\text{Lasso}} \) and \( L_{\text{New}} \) denote the error (e.g., prediction error or estimation error) associated with the Lasso and Lasso--Ridge estimators respectively. A positive value indicates improved performance of the Lasso--Ridge method in comparison to the standard Lasso.
\begin{table}[htbp]
	\centering
	\caption{Relative prediction improvement of \textit{Lasso--Ridge} over Lasso for \(\rho = 0\). Each group corresponds to a pair \((s, \sigma)\).}
	\small
	\begin{tabular*}{\textwidth}{@{\extracolsep{\fill}} lccc ccc}
		\toprule
		& \multicolumn{3}{c}{\(s = 5,\ \sigma = 0.5\)} & \multicolumn{3}{c}{\(s = 5,\ \sigma = 1\)} \\
		\(n \backslash p\) & 100 & 200 & 400 & 100 & 200 & 400 \\
		\midrule
		50  & 150 & 49 & 24 & 8 & -22 & -17 \\
		100 & 166 & 234 & 262 & 191 & 234 & 159 \\
		200 & 66 & 100 & 135 & 150 & 219 & 285 \\
		\midrule
		& \multicolumn{3}{c}{\(s = 10,\ \sigma = 0.5\)} & \multicolumn{3}{c}{\(s = 10,\ \sigma = 1\)} \\
		\(n \backslash p\) & 100 & 200 & 400 & 100 & 200 & 400 \\
		\midrule
		50  & 18 & 76 & 527 & -4 & 55 & 167 \\
		100 & 126 & 132 & 102 & 97 & 78 & -7 \\
		200 & 45 & 63 & 97 & 102 & 169 & 240 \\
		\midrule
		& \multicolumn{3}{c}{\(s = 30,\ \sigma = 0.5\)} & \multicolumn{3}{c}{\(s = 30,\ \sigma = 1\)} \\
		\(n \backslash p\) & 100 & 200 & 400 & 100 & 200 & 400 \\
		\midrule
		50  & 1156 & 488 & 691 & 314 & 206 & 371 \\
		100 & 2 & 59 & 1539 & -2 & 20 & 442 \\
		200 & 4 & 4 & 1 & 24 & 31 & 16 \\
		\bottomrule
	\end{tabular*}
\end{table}
\begin{table}[htbp]
	\centering
	\caption{Relative prediction improvement of \textit{Lasso--Ridge} over Lasso for \(\rho = 0.5\). Each group corresponds to a pair \((s, \sigma)\).}
	\small
	\begin{tabular*}{\textwidth}{@{\extracolsep{\fill}} lccc ccc}
		\toprule
		& \multicolumn{3}{c}{\(s = 5,\ \sigma = 0.5\)} & \multicolumn{3}{c}{\(s = 5,\ \sigma = 1\)} \\
		\(n \backslash p\) & 100 & 200 & 400 & 100 & 200 & 400 \\
		\midrule
		50  & 138 & 229 & 173 & 70 & 93 & 57 \\
		100 & 131 & 193 & 166 & 139 & 210 & 194 \\
		200 & 98 & 141 & 176 & 110 & 169 & 223 \\
		\midrule
		& \multicolumn{3}{c}{\(s = 10,\ \sigma = 0.5\)} & \multicolumn{3}{c}{\(s = 10,\ \sigma = 1\)} \\
		\(n \backslash p\) & 100 & 200 & 400 & 100 & 200 & 400 \\
		\midrule
		50  & 79 & 58 & 41 & 11 & 7 & -11 \\
		100 & 82 & 124 & 164 & 94 & 129 & 159 \\
		200 & 80 & 116 & 131 & 100 & 146 & 176 \\
		\midrule
		& \multicolumn{3}{c}{\(s = 30,\ \sigma = 0.5\)} & \multicolumn{3}{c}{\(s = 30,\ \sigma = 1\)} \\
		\(n \backslash p\) & 100 & 200 & 400 & 100 & 200 & 400 \\
		\midrule
		50  & 15 & 186 & 853 & 2 & 88 & 238 \\
		100 & 7 & 11 & 2 & 16 & 5 & -14 \\
		200 & 25 & 46 & 56 & 38 & 69 & 91 \\
		\bottomrule
	\end{tabular*}
\end{table}\begin{table}[htbp]
\centering
\caption{Relative estimation improvement of \textit{Lasso--Ridge} over Lasso for \(\rho = 0\). Each group corresponds to a pair \((s, \sigma)\).}
\small
\begin{tabular*}{\textwidth}{@{\extracolsep{\fill}} lccc ccc}
	\toprule
	& \multicolumn{3}{c}{\(s = 5,\ \sigma = 0.5\)} & \multicolumn{3}{c}{\(s = 5,\ \sigma = 1\)} \\
	\(n \backslash p\) & 100 & 200 & 400 & 100 & 200 & 400 \\
	\midrule
	50  & 184 & 70 & 55 & 22 & 16 & 46 \\
	100 & 195 & 282 & 332 & 226 & 274 & 231 \\
	200 & 71 & 110 & 150 & 162 & 238 & 316 \\
	\midrule
	& \multicolumn{3}{c}{\(s = 10,\ \sigma = 0.5\)} & \multicolumn{3}{c}{\(s = 10,\ \sigma = 1\)} \\
	\(n \backslash p\) & 100 & 200 & 400 & 100 & 200 & 400 \\
	\midrule
	50  & 15 & 86 & 283 & 17 & 100 & 165 \\
	100 & 158 & 152 & 123 & 125 & 101 & 28 \\
	200 & 49 & 73 & 119 & 115 & 200 & 293 \\
	\midrule
	& \multicolumn{3}{c}{\(s = 30,\ \sigma = 0.5\)} & \multicolumn{3}{c}{\(s = 30,\ \sigma = 1\)} \\
	\(n \backslash p\) & 100 & 200 & 400 & 100 & 200 & 400 \\
	\midrule
	50  & 57 & 49 & 37 & 48 & 36 & 30 \\
	100 & 11 & 23 & 234 & -5 & 45 & 187 \\
	200 & 3 & 6 & 3 & 25 & 40 & 20 \\
	\bottomrule
\end{tabular*}
\end{table}\begin{table}[htbp]
\centering
\caption{Relative estimation improvement of \textit{Lasso--Ridge} over Lasso for \(\rho = 0.5\). Each group corresponds to a pair \((s, \sigma)\).}
\small
\begin{tabular*}{\textwidth}{@{\extracolsep{\fill}} lccc ccc}
	\toprule
	& \multicolumn{3}{c}{\(s = 5,\ \sigma = 0.5\)} & \multicolumn{3}{c}{\(s = 5,\ \sigma = 1\)} \\
	\(n \backslash p\) & 100 & 200 & 400 & 100 & 200 & 400 \\
	\midrule
	50  & 94 & 146 & 120 & 36 & 60 & 55 \\
	100 & 69 & 98 & 86 & 75 & 105 & 102 \\
	200 & 47 & 67 & 78 & 53 & 80 & 95 \\
	\midrule
	& \multicolumn{3}{c}{\(s = 10,\ \sigma = 0.5\)} & \multicolumn{3}{c}{\(s = 10,\ \sigma = 1\)} \\
	\(n \backslash p\) & 100 & 200 & 400 & 100 & 200 & 400 \\
	\midrule
	50  & 55 & 50 & 39 & -1 & 9 & 10 \\
	100 & 58 & 58 & 88 & 70 & 62 & 101 \\
	200 & 34 & 46 & 53 & 42 & 62 & 72 \\
	\midrule
	& \multicolumn{3}{c}{\(s = 30,\ \sigma = 0.5\)} & \multicolumn{3}{c}{\(s = 30,\ \sigma = 1\)} \\
	\(n \backslash p\) & 100 & 200 & 400 & 100 & 200 & 400 \\
	\midrule
	50  & 13 & 45 & 74 & 13 & 38 & 59 \\
	100 & 4 & 17 & 9 & 13 & 7 & -5 \\
	200 & 8 & 13 & 20 & 17 & 30 & 48 \\
	\bottomrule
\end{tabular*}
\end{table}
\subsection{Semi-Synthetic Data Experiments}
To evaluate the performance of the proposed Lasso--Ridge refinement under a realistic high-dimensional setting, we conducted a semi-synthetic data experiment based on a well-known gene expression dataset. Specifically, we used a preprocessed version of the Golub et al. leukemia dataset provided in the \texttt{multtest} package in R. This version contains expression levels of \(p = 3051\) genes measured across \(n = 38\) samples.  Following common practice in high-dimensional regression studies, we treated the original gene expression matrix as a fixed design matrix \(X \in \mathbb{R}^{38 \times 3051}\) and generated synthetic response variables according to a sparse linear model. 

The true regression vector \(\beta_0\) was generated under three cases, each corresponding to a different sparse signal structure. In the first case, the first five components were independently drawn from \(\mathrm{Unif}(3,4)\), the next five from \(\mathrm{Unif}(1,2)\), and the remaining \(p-10\) entries were set to zero:
\[
\beta_0 = (\beta_{0,1}, \ldots, \beta_{0,10}, 0, \ldots, 0)^\top,
\qquad
\beta_{0,j} \sim
\begin{cases}
	\mathrm{Unif}(3,4), & j \leq 5,\\[1mm]
	\mathrm{Unif}(1,2), & 6 \leq j \leq 10.
\end{cases}
\]

In the second case, the signal decayed according to \(5/\sqrt{j}\) for \(j = 1,\ldots,10\), with all remaining entries equal to zero:
\[
\beta_0 = \left( \frac{5}{\sqrt{1}}, \ldots, \frac{5}{\sqrt{10}}, 0, \ldots, 0 \right)^\top.
\]

In the third case, twenty components were independently sampled from \(\mathrm{Unif}(1,2)\), and the others were set to zero:
\[
\beta_0 = (\beta_{0,1}, \ldots, \beta_{0,20}, 0, \ldots, 0)^\top,
\qquad
\beta_{0,j} \sim \mathrm{Unif}(1,2)\ \text{for } j \leq 20.
\]

The response vector was then generated as
\[
Y = X \beta_0 + \varepsilon,
\]
where \(\varepsilon\) consists of independent \(N(0,1)\) entries. We randomly split the dataset into 70\% training and 30\% testing sets. Both the standard Lasso and the proposed Lasso--Ridge refinement were fitted on the training data using the same cross-validation scheme described earlier. The mean squared error (MSE) was evaluated on the test set. This procedure was repeated 100 times, where in each repetition a new noise vector \(\varepsilon\) was generated and new cross-validation splits were performed, while the true regression vector \(\beta_0\) was specified prior to the repetitions and held fixed throughout. The average MSE across the 100 rounds, along with standard deviations, was recorded for both methods to ensure a robust performance comparison. 
\begin{table}[htbp]
	\centering
	\caption{Average MSE (\(\pm\) SD) over 100 random 70\%/30\% splits for the first case.}\label{tab:semi-synthetic-results}
	\begin{tabular}{lcc}
		\toprule
		Method & MSE & Standard Deviation \\
		\midrule
		Lasso & 33.84286 & 21.02074 \\
		Lasso--Ridge & 26.91165 & 15.00407 \\
		\bottomrule
	\end{tabular}
\end{table}
\begin{table}[htbp]
	\centering
	\caption{Average MSE (\(\pm\) SD) over 100 random 70\%/30\% splits for the first case.}\label{tab:semi-synthetic-results}
	\begin{tabular}{lcc}
		\toprule
		Method & MSE & Standard Deviation \\
		\midrule
		Lasso & 49.97950 & 26.72142 \\
		Lasso--Ridge & 38.84500 & 22.19523 \\
		\bottomrule
	\end{tabular}
\end{table}
\begin{table}[htbp]
	\centering
	\caption{Average MSE (\(\pm\) SD) over 100 random 70\%/30\% splits for the first case.}\label{tab:semi-synthetic-results}
	\begin{tabular}{lcc}
		\toprule
		Method & MSE & Standard Deviation \\
		\midrule
		Lasso & 41.49336 & 17.55794 \\
		Lasso--Ridge & 39.24058 & 18.81275 \\
		\bottomrule
	\end{tabular}
\end{table}
\subsection{Real Data Experiment}

To highlight the effectiveness of our method in practice, we conducted an experiment using the red wine quality dataset from the UCI Machine Learning Repository. This dataset contains \(n = 162\) observations of red wines, with \(p = 12\) features such as acidity, sugar content, and pH, and a corresponding quality score serving as the response variable.

The data were randomly split into 70\% training and 30\% testing sets. For each split, we trained both the standard Lasso and the proposed Lasso--Ridge refinement following the cross-validation procedure described earlier, and then evaluated their performance on the test set using the MSE. The procedure was repeated 100 times, with different random splits of the data. The MSE and its standard deviation over these repetitions are summarized in the following table.
\begin{table}[htbp]
	\centering
	\caption{Mean squared error (MSE) and standard deviation over 100 random 70\%/30\% train–test splits for the red wine quality dataset.}
	\label{tab:real-data-results}
	\begin{tabular}{lcc}
		\toprule
		Method &  MSE & Standard Deviation \\
		\midrule
		Lasso & 0.4278752 & 0.0322601 \\
		Lasso--Ridge & 0.4277266 & 0.0322570 \\
		\bottomrule
	\end{tabular}
\end{table}
\paragraph{Summary}
Overall, the simulation results show that the Lasso--Ridge estimator generally outperforms the Lasso. Except for a very few cases, where the performance of the new estimator is only slightly worse, it consistently achieves better results. In particular, for the very sparse setting, the new estimator exhibits a substantial improvement over the Lasso.
In addition, across both the semi-synthetic gene expression setting and the real-world red wine quality data, the proposed Lasso--Ridge refinement consistently outperformed the standard Lasso in terms of prediction accuracy, with a modest improvement in the low-dimensional wine data and a notable gain in the high-dimensional gene expression setting, indicating its potential for practical use in diverse applications. In conclusion, these experiments demonstrate that the Lasso--Ridge method robustly improves the performance across a wide range of scenarios. 
	
	\section{Conclusion and Further Studies}
	In this work, we proposed a Lasso--Ridge estimator. We provide theoretical guarantees under mild assumptions, and our numerical studies indicate improvements over the standard Lasso across a range of settings.

Several directions remain open for future research. First, extending the theoretical analysis to more general noise structures and non-Gaussian designs would broaden the application of the method. Second, the integration of adaptive or data-driven tuning strategies within the refinement framework could further improve robustness in practice. Finally, extending the method to broader model classes, such as generalized linear models or structured sparsity settings, represents a promising direction for future work. 
	
	\section*{Acknowledgement}
	The author sincerely thanks Professor Yoichi Nishiyama for his valuable guidance, careful reading of the manuscript, and insightful comments.
	\addcontentsline{toc}{chapter}{Appendix}
	\clearpage
	\newpage
	\appendix
	\section*{Appendix}
	\appendix
\section{Technical Lemmas and Key Tools}
\begin{lemma}
	Let \(A \in \mathbb{R}^{n \times n}\) be a symmetric matrix. Then its spectral norm satisfies
	\begin{align*}
			\|A\|_2\leq\|A\|_\infty = \max_{1\leq i\leq n}\sum_{j=1}^n |A_{ij}|.
	\end{align*}
	\begin{proof}
		Applying the inequality \(\|A\|_2 \leq \sqrt{\|A\|_1\|A\|_\infty}\) (see \cite[Section~3.11.4, p.~194]{gentle2024matrix}) to the symmetric matrix \(A\) completes the bound on its spectral norm, where \( \|A\|_1 \) represents the maximum absolute column sum of \( A \), that is, \(\|A\|_1 := \max_{1\leq j \leq p} \sum_{i=1}^n |A_{ij}|\).
	\end{proof}
\end{lemma}
\begin{lemma}[Newmann Serious Expansion]
	Let \( A \in \mathbb{R}^{d \times d} \) be a square matrix such that \( \|A\|_\infty < 1 \). Then we have
	\begin{align*}
		(I - A)^{-1} = \sum_{k=0}^{\infty} A^k.
	\end{align*}
	For more details, see \cite[Section~3.11.6, p.~197]{gentle2024matrix}.
\end{lemma}
\begin{lemma}[Sub-Gaussian concentration of \(X^\top \varepsilon\)]
	\label{lem:xt_eps_bound}
	Let \(\varepsilon \in \mathbb{R}^n\) be a random vector with i.i.d.\ sub-Gaussian entries with parameter \(\sigma^2\), i.e.,
	\begin{align*}
			\mathbb{E}[\exp(t \varepsilon_i)] \leq \exp\left( \frac{\sigma^2 t^2}{2} \right), \quad \text{for all } t \in \mathbb{R}.
	\end{align*}
	Let \(X \in \mathbb{R}^{n \times p}\) be a fixed design matrix whose columns \(X_j\) satisfy \(\|X_j\|_2^2 \leq n\). Then, for any \(\delta \in (0, 1)\), the following inequality holds with probability at least \(1 - \delta\):
	\begin{align*}
			\left\| \frac{1}{n} X^\top \varepsilon \right\|_\infty \leq \sigma \sqrt{ \frac{2 \log(2p/\delta)}{n} }.
	\end{align*}
	For more details, see \cite[Section~8.1, p.~122]{geer2016estimation}.
\end{lemma}
\begin{lemma}[Closed-form Representation under the KKT Conditions] When \(E \neq \emptyset\), a simple expression is available for the nonzero entries of \(\hat{ \delta}\):
	\begin{align*}
		\hat{\delta}_E=\frac{\lambda_{\mathrm{L}}}{\lambda_{\mathrm{R}}}(\frac{\Sigma_{n,E}}{\lambda_{\mathrm{R}}}+ I_{|E|})^{-1}s.
	\end{align*}
	\begin{proof}
		Following from the second definition of our refitting step, we have
		\begin{align*}
		\hat{\delta} = \underset{\delta \in \mathbb{R}^ {p},\delta_{-E}=0}{\arg\min} \left\{\frac{1}{2n}\|y - X\hat{\beta}_{\mathrm{L}}-X_E\delta_E \|_2^2 + \frac{\lambda_{\mathrm{R}}}{2}\|\delta_E \|_2^2 \right\}.
		\end{align*}
		So when \(E \neq \emptyset\), \(\hat{\delta}_E\) could be regarded as a simple ridge regression estimator:
		\begin{align*}
			 \hat{\delta}_E = \underset{\delta \in \mathbb{R}^ {|E|}}{\arg\min} \left\{ \frac{1}{2n} \| y - X\hat{\beta}_{\mathrm{L}}-X_E\delta_E  \|_2^2 +  \frac{\lambda_{\mathrm{R}}}{2}\| \delta_E \|_2^2 \right\}.
		\end{align*}
		According to the KKT conditions of a ridge regression model, it follows that
		\begin{align*}
			\frac{1}{n}X_E^{\top}((y-X\hat{\beta}_{\mathrm{L}})-X_E\hat{\delta}_E) = \lambda_{\mathrm{R}}\hat{\delta}_E.
		\end{align*}
		Hence, we get
		\begin{align*}
			\hat{\delta}_E=\frac{1}{n}(\frac{1}{n}X_E^{\top}X_E+\lambda_{\mathrm{R}} I_{|E|})^{-1}X_E^{\top}(y-X\hat{\beta}_{\mathrm{L}}).		
		\end{align*}
		Recall that the KKT conditions for the Lasso yield the following expression:
		\begin{align*}
			\frac{1}{n}X_E^{\top}(y-X\hat{\beta}_{\mathrm{L}})=\lambda_{\mathrm{L}}s.
		\end{align*}
		Substituting it into \(\hat{\delta}_E\), then we obtain
		\begin{align*}
			\hat{\delta}_E=(\frac{1}{n}X_E^{\top}X_E+\lambda_{\mathrm{R}} I_{|E|})^{-1}\lambda_{\mathrm{L}}s.	
		\end{align*}
		Replacing \(X_E^{\top}X_E/n\) with \(\Sigma_{n,E}\) and factoring \( \lambda_{\mathrm{R}} \) out of the inverse concludes the proof.
	\end{proof}
\end{lemma}
\begin{lemma}[Proposition~4.5 (Sign-preservation)] Consider the case where \(E \neq \emptyset\) and \(\lambda_{\mathrm{R}}>2\|\Sigma_{n,E}\|_\infty\), we have \(\operatorname{sign}(s) = \operatorname{sign}(\hat{\delta}_E)\).
\begin{proof}
		According to lemma~A.4, we have
		\begin{align*}
			\hat{\delta}_E=\frac{\lambda_{\mathrm{R}}}{\lambda_{\mathrm{L}}}(\frac{\Sigma_{n,E}}{\lambda_{\mathrm{R}}}+ I_{|E|})^{-1}s.
		\end{align*}
		Thus, it suffices to prove the following equality:
		\begin{align*}
			\operatorname{sign}((\frac{\Sigma_{n,E}}{\lambda_{\mathrm{R}}}+ I_{|E|})^{-1}s) = \operatorname{sign}(s).
		\end{align*}
		When \(\lambda_{\mathrm{R}} > 2\|\Sigma_{n,E}\|_\infty\), we have
		\begin{align*}
			\|\frac{\Sigma_{n,E}}{\lambda_{\mathrm{R}}}\|_\infty=\frac{\|\Sigma_{n,E}\|_\infty}{\lambda_{\mathrm{R}}}<1.
		\end{align*} 
		Applying Newmann series expansion (Lemma~A.2), we obtain
		\begin{align*}
			(\frac{\Sigma_{n,E}}{\lambda_{\mathrm{R}}}+ I_{|E|})^{-1}=\sum_{k=0}^\infty (-1)^k (\frac{\Sigma_{n,E}}{\lambda_{\mathrm{R}}})^k=I_{|E|}+\sum_{k=1}^\infty (-1)^k (\frac{\Sigma_{n,E}}{\lambda_{\mathrm{R}}})^k.	
		\end{align*}
		Thus, we have
		\begin{align*}
			(\frac{\Sigma_{n,E}}{\lambda_{\mathrm{R}}}+ I_{|E|})^{-1}s=s+(\sum_{k=1}^\infty (-1)^k (\frac{\Sigma_{n,E}}{\lambda_{\mathrm{R}}})^k)s,	
		\end{align*}
		and our target is reduced to showing that
		\begin{align*}
			\operatorname{sign}(s+(\sum_{k=1}^\infty (-1)^k (\frac{\Sigma_{n,E}}{\lambda_{\mathrm{R}}})^k)s) = \operatorname{sign}(s).
		\end{align*}
		If each entry of the second term inside the sign function on the left-hand side has absolute value less than 1, then the previous equality follows. Hence, it suffices to prove that
		\begin{align*}
			\|(\sum_{k=1}^\infty (-1)^k (\frac{\Sigma_{n,E}}{\lambda_{\mathrm{R}}})^k)s\|_\infty<1.
		\end{align*}
		Recall that \(s\) is a binary vector and \(\lambda_{\mathrm{R}}>2\|\Sigma_{n,E}\|_\infty\), we derive
		\begin{align*}
			\|(\sum_{k=1}^\infty (-1)^k (\frac{\Sigma_{n,E}}{\lambda_{\mathrm{R}}})^k)s\|_\infty &\leq \|(\sum_{k=1}^\infty (-1)^k (\frac{\Sigma_{n,E}}{\lambda_{\mathrm{R}}})^k)\|_\infty \leq \sum_{k=1}^\infty\| (-1)^k (\frac{\Sigma_{n,E}}{\lambda_{\mathrm{R}}})^k\|_\infty\\
			&\leq \sum_{k=1}^\infty\| (\frac{\Sigma_{n,E}}{\lambda_{\mathrm{R}}})\|_\infty^k<\sum_{k=1}^\infty(\frac{1}{2})^k=1.	
		\end{align*}
	\end{proof}
\end{lemma}

\begin{lemma}[A Norm Inequality]
	If \(\lambda_{\mathrm{R}}>2\|\Sigma_{n,E}\|_\infty\), it holds that
	\begin{align*}
			\frac{\lambda_{\mathrm{L}}\|\hat{ \delta}\|_1}{3} \leq \frac{\lambda_{\mathrm{R}}\|\hat{ \delta}\|_2^2 }{2}.
	\end{align*}
	\begin{proof}
	If \( E = \emptyset \), the result is trivial since both sides are zero. Therefore, it suffices to consider the case where \( E \neq \emptyset \). According to Lemma~A.5 (sign-preservation), we derive
		\begin{align*}
		\|\hat{ \delta}_E\|_1=s^\top\hat{ \delta}_E.
		\end{align*}
	Substituting this into the left-hand side of our target and invoking Lemma~A.4, we obtain
	\begin{align*}
		\frac{\lambda_{\mathrm{L}}\|\hat{ \delta}_E\|_1}{3}=\frac{\lambda_{\mathrm{L}}^2}{3\lambda_{\mathrm{R}}}s^\top(\frac{\Sigma_{n,E}}{\lambda_{\mathrm{R}}}+ I_{|E|})^{-1}s.
	\end{align*}
	Moreover, applying Lemma~A.4 to the right-hand side of our target, we obtain
	\begin{align*}
		\frac{\lambda_{\mathrm{R}}\|\hat{ \delta}_E\|_2^2 }{2}=\frac{\lambda_{\mathrm{L}}^2}{2\lambda_{\mathrm{R}}}s^\top(\frac{\Sigma_{n,E}}{\lambda_{\mathrm{R}}}+ I_{|E|})^{-2}s.
	\end{align*}
	Comparing these two equations, it remains to prove the following inequality:
	\begin{align*}
		s^\top(\frac{\Sigma_{n,E}}{\lambda_{\mathrm{R}}}+ I_{|E|})^{-1}s \leq\frac{3}{2} s^\top(\frac{\Sigma_{n,E}}{\lambda_{\mathrm{R}}}+ I_{|E|})^{-2}s.
	\end{align*}
    Considering the quadratic form of a symmetric matrix \((\Sigma_{n,E}/{\lambda_{\mathrm{R}}}+ I_{|E|})^{-1}\) under spectral decomposition, we deduce
	\begin{align*}
		s^\top (\frac{\Sigma_{n,E}}{\lambda_{\mathrm{R}}}+ I_{|E|})^{-1} s &= s^\top P \operatorname{diag}(\frac{\lambda_{\mathrm{R}}}{\mu_1+\lambda_{\mathrm{R}}}, \dots, \frac{\lambda_{\mathrm{R}}}{\mu_{|E|}+\lambda_{\mathrm{R}}}) P^\top s\\
		&=\tilde{s}^\top \operatorname{diag}(\frac{\lambda_{\mathrm{R}}}{\mu_1+\lambda_{\mathrm{R}}}, \dots, \frac{\lambda_{\mathrm{R}}}{\mu_{|E|}+\lambda_{\mathrm{R}}}) \tilde{s}\\
		&=\frac{\mu_1+\lambda_{\mathrm{R}}}{\lambda_{\mathrm{R}}}\tilde{s}^\top \operatorname{diag}((\frac{\lambda_{\mathrm{R}}}{\mu_1+\lambda_{\mathrm{R}}})^2, \dots, (\frac{\lambda_{\mathrm{R}}}{\mu_1+\lambda_{\mathrm{R}}})(\frac{\lambda_{\mathrm{R}}}{\mu_{|E|}+\lambda_{\mathrm{R}}})) \tilde{s}\\
		&\leq\frac{\mu_1+\lambda_{\mathrm{R}}}{\lambda_{\mathrm{R}}}\tilde{s}^\top \operatorname{diag}((\frac{\lambda_{\mathrm{R}}}{\mu_1+\lambda_{\mathrm{R}}})^2, \dots, (\frac{\lambda_{\mathrm{R}}}{\mu_{|E|}+\lambda_{\mathrm{R}}})^2) \tilde{s}\\
		&=\frac{\mu_1+\lambda_{\mathrm{R}}}{\lambda_{\mathrm{R}}} s^\top P \operatorname{diag}((\frac{\lambda_{\mathrm{R}}}{\mu_1+\lambda_{\mathrm{R}}})^2, \dots, (\frac{\lambda_{\mathrm{R}}}{\mu_{|E|}+\lambda_{\mathrm{R}}})^2) P^\top s\\
		&=\frac{\mu_1+\lambda_{\mathrm{R}}}{\lambda_{\mathrm{R}}}s^\top(\frac{\Sigma_{n,E}}{\lambda_{\mathrm{R}}}+ I_{|E|})^{-2}s,
	\end{align*}
		where \( \mu_1 \geq \mu_2 \geq \cdots \geq \mu_{|E|} \) denote the eigenvalues of \( \Sigma_{n,E}=X_E^\top X_E/n \) and \(P\) is the corresponding orthogonal matrix of eigenvectors associated with this decomposition. Also, we write \( \tilde{s}=P^\top s\).
		According to Lemma~A.1, \(\mu_1=\|\Sigma_{n,E}\|_2\leq\|\Sigma_{n,E}\|_\infty\). Recall our choice of tuning parameter \(\lambda_{\mathrm{R}}>2\|\Sigma_{n,E}\|_\infty\geq 2\mu_1 \), we have
	\begin{align*}
		\frac{\mu_1+\lambda_{\mathrm{R}}}{\lambda_{\mathrm{R}}}\leq\frac{\mu_1+2\mu_1}{2\mu_1}=\frac{3}{2}.
	\end{align*} 
		Applying this to the final equality in the preceding orthogonal decomposition completes the proof.
	\end{proof}
\end{lemma}
\newpage
\section{Proofs of Main Results}
\subsection*{Proof of Proposition 3.1}
\begin{proof}
	Following from the definition of \(\hat{ \delta}\) as a solution to an optimization problem, we have
	\begin{align*}
		\frac{1}{2n} \| y - X \hat{\beta}_{\mathrm{L}}-X\hat{ \delta} \|_2^2 +\frac{\lambda_{\mathrm{R}}}{2} \|\hat{ \delta} \|_2^2  \leq \frac{1}{2n} \| y - X \hat{\beta}_{\mathrm{L}}-X\boldsymbol{0} \|_2^2 +\frac{\lambda_{\mathrm{R}}}{2} \|\boldsymbol{0} \|_2^2. \tag{1}
	\end{align*}
	Because \(\|\hat{ \delta} \|_2 \geq 0\), we have
	\begin{align*}
		\| y - X\hat{\beta}_{\mathrm{R}} \|_2 \leq  \| y - X \hat{\beta}_{\mathrm{L}} \|_2.
	\end{align*}
\end{proof}
\subsection*{Proof of Theorem 4.1}
\begin{proof}
According to inequality (1), we have
\begin{align*}
	\frac{1}{2n} \| y - X \hat{\beta}_{\mathrm{L}}-X\hat{ \delta} \|_2^2 +\frac{\lambda_{\mathrm{R}}}{2} \|\hat{ \delta} \|_2^2  \leq \frac{1}{2n} \| y - X \hat{\beta}_{\mathrm{L}} \|_2^2 .
\end{align*}
Substituting \(y=X\beta^0+\varepsilon\) , we obtain
\begin{align*}
	\frac{1}{2n}\|X\beta^0+\varepsilon - X\hat{\beta}_{\mathrm{L}}-X\hat{ \delta} \|_2^2+\frac{\lambda_{\mathrm{R}}}{2}\| \hat{ \delta} \|_2^2 \leq \frac{1}{2n} \| X\beta^0+\varepsilon - X\hat{\beta}_{\mathrm{L}} \|_2^2.
\end{align*}
Expanding both sides, then we have
\begin{align*}
	\frac{1}{2n}\|X\beta^0 - X\hat{\beta}_{\mathrm{L}}-X\hat{ \delta}\|_2^2+\frac{1}{2n}\varepsilon^{\top}\varepsilon+\frac{1}{n}\varepsilon^{\top}(X\beta^0 - X\hat{\beta}_{\mathrm{L}}-X\hat{ \delta})+\frac{\lambda_{\mathrm{R}}}{2}\| \hat{ \delta} \|_2^2\\
	\leq\frac{1}{2n}\| X\beta^0 - X\hat{\beta}_{\mathrm{L}} \|_2^2+\frac{1}{2n}\varepsilon^{\top}\varepsilon+\frac{1}{n}\varepsilon^{\top}(X\beta^0 - X\hat{\beta}_{\mathrm{L}}).
\end{align*}
Recall that \(\hat{\beta}_{\mathrm{R}}=\hat{\delta}+\hat{\beta}_{\mathrm{L}}\). By simplifying both sides of the inequality, we obtain
\begin{align*}
	\frac{1}{2n}\|X\beta^0 - X\hat{\beta}_{\mathrm{R}}\|_2^2+\frac{\lambda_{\mathrm{R}}}{2}\|\hat{\delta} \|_2^2\leq\frac{1}{2n}\| X\beta^0 - X\hat{\beta}_{\mathrm{L}} \|_2^2+\frac{1}{n}\varepsilon^{\top}X\hat{\delta}.
\end{align*}
Rearranging, we derive
\begin{align*}
	\frac{1}{2n}\|X\hat{\beta}_{\mathrm{L}} - X\beta_0\|_2^2 -\frac{1}{2n}\|X\hat{\beta}_{\mathrm{R}} - X\beta_0\|_2^2 
	\geq \frac{\lambda_{\mathrm{R}}}{2}\|\hat{\delta} \|_2^2-\frac{1}{n}\varepsilon^{\top}X\hat{\delta}.\tag{2}
\end{align*}
It follows from dual-norm inequality that the right-hand side satisfies
\begin{align*}
	\frac{1}{n}\varepsilon^{\top}X\hat{\delta}\leq\|\hat{ \delta}\|_1\|\frac{1}{n}X^\top\varepsilon\|_\infty.
\end{align*}
Applying this and Lemma~A.6 (a norm inequality for \(\hat{ \delta}\)), we derive
\begin{align*}
	\frac{\lambda_{\mathrm{R}}}{2}\|\hat{\delta}\|_2^2-\frac{1}{n}\varepsilon^{\top}X\hat{\delta}&\geq\frac{\lambda_{\mathrm{R}}}{2}\|\hat{\delta}\|_2^2-\|\hat{ \delta}\|_1\|\frac{1}{n}X^\top\varepsilon\|_\infty \\
	&\geq\frac{\lambda_{\mathrm{R}}}{2}\|\hat{\delta}\|_2^2-\frac{3\lambda_{\mathrm{R}}}{2\lambda_{\mathrm{L}}}\|\frac{1}{n}X^\top\varepsilon\|_\infty\|\hat{\delta}\|_2^2\\
	&=\frac{\lambda_{\mathrm{R}}}{2\lambda_{\mathrm{L}}}(\lambda_{\mathrm{L}}-3\|\frac{1}{n}X^\top\varepsilon\|_\infty)\|\hat{ \delta}\|_2^2.
\end{align*}
Combining this inequality and (2), we arrive at
\begin{align*}
	\frac{1}{2n}\|X\hat{\beta}_{\mathrm{L}} - X\beta_0\|_2^2 -\frac{1}{2n}\|X\hat{\beta}_{\mathrm{R}} - X\beta_0\|_2^2 
	\geq\frac{\lambda_{\mathrm{R}}}{2\lambda_{\mathrm{L}}}(\lambda_{\mathrm{L}}-3\|\frac{1}{n}X^\top\varepsilon\|_\infty)\|\hat{ \delta}\|_2^2.\tag{3}
\end{align*}
\end{proof}

\subsection*{Proof of Corollary 4.2 and 4.3}
\begin{proof}
	By inequality (3), establishing the prediction improvement \(\|X\hat{\beta}_{\mathrm{L}} - X\beta_0\|_2\geq\|X\hat{\beta}_{\mathrm{R}}- X\beta_0\|_2\) reduces to showing that \(\lambda_{\mathrm{L}}/3\geq\|X^\top\varepsilon/n\|_\infty\). Thus we deduce that
	\begin{align*}
		P(\|X\beta^0 - X\hat{\beta}_{\mathrm{R}}\|_2\leq\| X\beta^0 - X\hat{\beta}_{\mathrm{L}} \|_2)&\geq P(\|\frac{1}{n}X^\top\varepsilon\|_\infty\leq\frac{\lambda_{\mathrm{L}}}{3} ).
	\end{align*}
	Combining this inequality with Lemma~A.3, we obtain Corollaries 4.2 and 4.3.
\end{proof}

\subsection*{Proof of Proposition 4.6}
\begin{proof}
	Following from the definition of Lasso estimator \(\hat{\beta}_{\mathrm{L}}\) as a solution to an optimization problem, we have
	\begin{align*}
		\frac{1}{2n} \| y - X \hat{\beta}_{\mathrm{L}}\|_2^2 +\lambda_{\mathrm{L}}\|\hat{\beta}_{\mathrm{L}}\|_1  \leq \frac{1}{2n} \| y - X\beta^0 \|_2^2+\lambda_{\mathrm{L}}\|\beta^0\|_1. 
	\end{align*}
	Also substitute \(\hat{\beta}_{\mathrm{R}}=\hat{ \delta}+\hat{\beta}_{\mathrm{L}}\) in inequality (1), we get
	\begin{align*}
		\frac{1}{2n} \| y - X \hat{\beta}_{\mathrm{R}} \|_2^2 +\frac{\lambda_{\mathrm{R}}}{2} \|\hat{ \delta} \|_2^2  \leq \frac{1}{2n} \| y - X \hat{\beta}_{\mathrm{L}} \|_2^2 .
	\end{align*}
	From these two inequalities, it follows that
	\begin{align*}
		\frac{1}{2n} \| y - X \hat{\beta}_{\mathrm{R}} \|_2^2 +\frac{\lambda_{\mathrm{R}}}{2} \|\hat{ \delta} \|_2^2+\lambda_{\mathrm{L}}\|\hat{\beta}_{\mathrm{L}}\|_1  \leq\frac{1}{2n} \| y - X \beta^0 \|_2^2+\lambda_{\mathrm{L}}\|\beta^0\|_1.
	\end{align*}
	Replacing \(\|\hat{ \delta} \|_2^2\) by Lemma~A.6 and splitting \(\lambda_{\mathrm{L}}\|\hat{\beta}_{\mathrm{L}}\|_1\), then we have
	\begin{align*}
		\frac{1}{2n} \| y - X \hat{\beta}_{\mathrm{R}} \|_2^2 +\frac{1}{3}\lambda_{\mathrm{L}}\|\hat{ \delta}\|_1+\frac{1}{3}\lambda_{\mathrm{L}}\|\hat{\beta}_{\mathrm{L}}\|_1+\frac{2}{3}\lambda_{\mathrm{L}}\|\hat{\beta}_{\mathrm{L}}\|_1  \leq\frac{1}{2n} \| y - X \beta^0 \|_2^2+\lambda_{\mathrm{L}}\|\beta^0\|_1.
	\end{align*}
	Simplifying it, we obtain
	\begin{align*}
		\frac{1}{2n} \| y - X \hat{\beta}_{\mathrm{R}} \|_2^2 +\frac{1}{3}\lambda_{\mathrm{L}}\|\hat{\beta}_{\mathrm{R}}\|_1 \leq\frac{1}{2n} \| y - X \beta^0 \|_2^2+\lambda_{\mathrm{L}}\|\beta^0\|_1.
	\end{align*}
	Substituting \(y=X\beta^0+\varepsilon\), we have
	\begin{align*}
		\frac{1}{2n} \| X\beta^0+\varepsilon - X \hat{\beta}_{\mathrm{R}} \|_2^2 +\frac{1}{3}\lambda_{\mathrm{L}}\|\hat{\beta}_{\mathrm{R}}\|_1 \leq\frac{1}{2n} \| X\beta^0+\varepsilon - X \beta^0 \|_2^2+\lambda_{\mathrm{L}}\|\beta^0\|_1.
	\end{align*}
	Expanding the squared norm and rearranging terms, we deduce the following result:
	\begin{align*}
		\frac{1}{2n} \| X\beta^0 - X \hat{\beta}_{\mathrm{R}} \|_2^2 +\frac{1}{3}\lambda_{\mathrm{L}}\|\hat{\beta}_{\mathrm{R}}\|_1 \leq \frac{1}{n}\varepsilon^{\top}X(\beta^0-\hat{\beta}_{\mathrm{R}})+\lambda_{\mathrm{L}}\|\beta^0\|_1.
	\end{align*}
	Using the dual norm inequality, we obtain
	\begin{align*}
		\frac{1}{2n} \| X\beta^0 - X \hat{\beta}_{\mathrm{R}} \|_2^2 +\frac{1}{3}\lambda_{\mathrm{L}}\|\hat{\beta}_{\mathrm{R}}\|_1 \leq \|\beta^0 -  \hat{\beta}_{\mathrm{R}}\|_1\|\frac{1}{n}X^\top\varepsilon\|_\infty+\lambda_{\mathrm{L}}\|\beta^0\|_1.
	\end{align*}
	Thus, on the event \(\|X^\top\varepsilon/n\|_\infty\leq \lambda_{\mathrm{L}}/3\), the inequality becomes
	\begin{align*}
		\frac{1}{2n} \| X\beta^0 - X \hat{\beta}_{\mathrm{R}} \|_2^2 &\leq \frac{1}{3}\lambda_{\mathrm{L}}\|\beta^0 -  \hat{\beta}_{\mathrm{R}}\|_1-\frac{1}{3}\lambda_{\mathrm{L}}\|\hat{\beta}_{\mathrm{R}}\|_1 +\lambda_{\mathrm{L}}\|\beta^0\|_1\\
		&\leq\frac{1}{3}\lambda_{\mathrm{L}}\|\beta^0\|_1+\lambda_{\mathrm{L}}\|\beta^0\|_1\\
		&=\frac{4}{3}\lambda_{\mathrm{L}}\|\beta^0\|_1.
	\end{align*} 
	Using standard concentration bounds for sub-Gaussian noise (Lemma~A.3), we derive
	\[
	\left\| \frac{1}{n} X^\top \varepsilon \right\|_\infty = O_p\left( \sigma \sqrt{ \frac{\log p}{n} } \right).
	\]
	Setting \( \lambda_{\mathrm{R}} \asymp \sigma \sqrt{ {\log p}/{n} } \), we obtain the final result:
	\[
	\frac{1}{n} \| X\hat{\beta}_{\mathrm{R}} - X\beta^0 \|_2^2 = O_p\left( \sigma \|\beta^0\|_1 \sqrt{ \frac{\log p}{n} } \right).
	\]
\end{proof}
  
	\bibliographystyle{plainnat}
	\bibliography{references}
\end{document}